# Calculation of Release Adiabats and Shock Impedance Matching

A Kerley Technical Services Research Report

Gerald I. Kerley, Consultant

March 2008

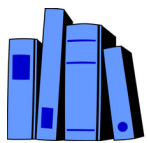

# Calculation of Release Adiabats and Shock Impedance Matching

## ABSTRACT

In the analysis of impedance-match experiments, the release adiabat of the standard material is often approximated by reflecting the Hugoniot in the pressure-particle velocity ($P$-$u_P$) plane. In cases where it has been tested experimentally, this reflected shock approximation (RSA) has been found to be fairly accurate at pressures as high as 100 GPa. The success of the RSA is usually attributed to the lack of thermal effects, i.e., that the Hugoniot and release adiabat are nearly identical in the pressure-density ($P$-$\rho$) plane. This report demonstrates that this explanation is not correct. The success of the RSA does not arise from the *absence* of thermal effects—it arises *because* of thermal effects. When the Hugoniot and adiabat are identical in the $P$-$\rho$ plane, the adiabat lies *below* the reflected Hugoniot when mapped into the $P$-$u_P$ plane. Material strength effects also cause the exact adiabat to lie *below* the RSA. A thermal offset between the Hugoniot and the adiabat compensates for these two effects, so that the RSA turns out to give good results for many materials, even at high pressures. However, this fortuitous cancellation of errors does not occur in all cases. This report shows that the RSA is not accurate for two "soft" materials, Teflon and PMMA, and for a high-strength material, tungsten. The issues discussed here apply to use of the RSA at *low* pressures, not to very high pressures, where it is already well-known to be inaccurate.

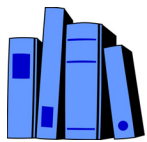

# CONTENTS

# FIGURES

# SYMBOLS AND UNITS

| | |
|---|---|
| $\rho$ | density [$g/cm^3$] |
| $P$ | pressure [$GPa$] |
| $U_S$ | shock velocity [$km/s$] |
| $u_P$ | particle velocity [$km/s$] |
| $u_R$ | particle velocity on release adiabat [$km/s$] |
| $\rho_0$ | density of solid at zero pressure and temperature [$g/cm^3$] |
| $C_0$ | sound velocity at zero pressure [$km/s$] |
| $\beta_0$ | bulk modulus of solid at zero pressure and temperature [$GPa$] |
| $\Gamma$ | Grüneisen function (used in Mie-Grüneisen EOS) |

# PREFACE

Several years ago, Craig Doolittle (Applied Research Associates, Albuquerque, NM) brought an interesting fact to my attention: his hydrocode simulation of a release wave experiment in aluminized-PTFE did not agree with predictions made using the reflected shock approximation (RSA) for the release adiabat. The shock pressure in this experiment was rather low, so that shock heating would not be expected to cause significant deviations from the RSA. Moreover, the observed deviations were the *opposite* of what would be expected from thermal effects.

My investigation into this matter revealed that the hydrocode results *were correct* and that the RSA did not give accurate results for this case, despite the low shock pressure and temperature. I later made a more systematic study of the conditions under which the RSA could fail at low shock pressures and temperatures. I also investigated the effects of material strength on the release adiabat. Some of these results were presented in my EOS course in Roanoke, VA, in June, 2005. They are also discussed briefly in the "Basic Concepts" tutorial on my website.

I have prepared this more detailed report because these issues do not appear to be widely known within the shock wave community. Deviations from the RSA at *high* shock pressures are discussed in the literature. However, I have not found any discussion of deviations from the RSA at *low* shock pressures—not in the classic 1958 paper of Rice, McQueen, and Walsh [1], or any of the other reviews in my possession.

I am grateful to Craig for bringing this matter to my attention in the first place.

**Added note, 08/05/09**: I have recently learned that problems with the RSA were first discussed by Julius Enig, almost 50 years ago, in a report [NOLTR 62-149 (1962), available from the internet], and two papers [JAP 34, 746 (1963); 35, 465 (1964)]. Enig's approach was different from mine but reached many of the same conclusions. He showed that the RSA, which he called the mirror-image assumption, could be used to derive a complete EOS for the material and that the results were inconsistent with the Mie-Grüneisen EOS. He pointed out that use of the RSA could lead to errors in the analysis of shock experiments. His ideas and his warning have been ignored for a half-century.

I thank Enig's former colleague, Kenneth L. Moore, now at Raytheon Missiles Systems Company, Tucson, AZ, for bringing this information to my attention.

# 1. INTRODUCTION

## 1.1 Background

An understanding of the unloading behavior of shocked materials is important for a variety of applications. The temperature on a release adiabat, or isentrope, is higher than on the Hugoniot, at the same density; this thermal offset can lead to interesting results, particularly at high shock pressures. The unloading behavior can also be affected by shock-induced phase changes and chemical reactions. Measurements of release adiabats can give information about materials that cannot be discovered solely from studies of the principal Hugoniot.

Shock waves are eventually overtaken by rarefaction waves in all materials. As the rarefaction degrades the pressure at the shock front, the material expands isentropically, while the particle velocity increases. Hence the release behavior affects shock propagation phenomena in all but the simplest problems. In hydrocode simulations, it is prudent to use equations of state (EOS) that have a realistic description of the release behavior as well as the Hugoniot.

The release curve is also required for analysis of certain experiments designed to measure the Hugoniot. Three important cases are:
- when the free surface velocity is used to determine the particle velocity of either a sample or a standard material;
- in an impedance matching[1] (IM) experiment, when the sample is initially in contact with a standard material having higher impedance; or
- when acceleration of an impactor leaves it in a heated, distended state.

The present report is motivated primarily by consideration of the first two cases, in which the release adiabat is frequently estimated by what I will call the "reflected shock approximation" (RSA). The RSA assumes that the release adiabat is a mirror image of the Hugoniot in the pressure-particle velocity plane, as illustrated below. The RSA is sometimes claimed to be reasonably accurate at pressures below 100 GPa; in practice, it is often used at even higher pressures.

It is well known that the RSA breaks down at high shock pressures, where thermal effects are important. In this report, I will show that the RSA can break down at very *low* pressures—*even when the Hugoniot and release adiabat are identical in the pressure-density plane.*

---

1. The shock impedance is given by $\rho_0 U_S$, where $\rho_0$ is the initial density and $U_S$ is the shock velocity at a given particle velocity $u_P$. Some authors prefer to use the term impedance *mis*-matching, since the sample and standard have different impedances. The abbreviation IM works in both cases.

## 1.2 Conventional Wisdom

First let us review the main concepts as they are currently understood throughout the shock-wave literature.

Figure 1 shows the Hugoniot (red curve) and a release adiabat (blue curve) for aluminum in the pressure-particle velocity ($P$-$u_P$) plane. Both curves were calculated from my theoretical EOS, discussed in Ref. [2], using the EOSPro code [3]. (For simplicity, material strength was not included in these calculations; strength effects will be discussed in Sec. 4.)

The release adiabat, which corresponds to an initial shock pressure of 30 GPa, was calculated in the pressure-density ($P$-$\rho$) plane, using the theoretical EOS, then mapped into the $P$-$u_P$ plane using the well-known Riemann integral [1]: Given an initial shock state with pressure $P_{H1}$, density $\rho_1$, and particle velocity $u_{P1}$, the rarefaction velocity $u_R$ (particle velocity along the adiabat) is given by

$$u_R = u_{P1} - \int_{\rho_1}^{\rho} C_S \, \rho^{-1} d\rho, \qquad (1)$$

where $C_S$ is the sound speed, related to the rarefaction pressure $P_R$ by

$$C_S = \sqrt{(\partial P/\partial \rho)_S} = \sqrt{dP_R/d\rho}. \qquad (2)$$

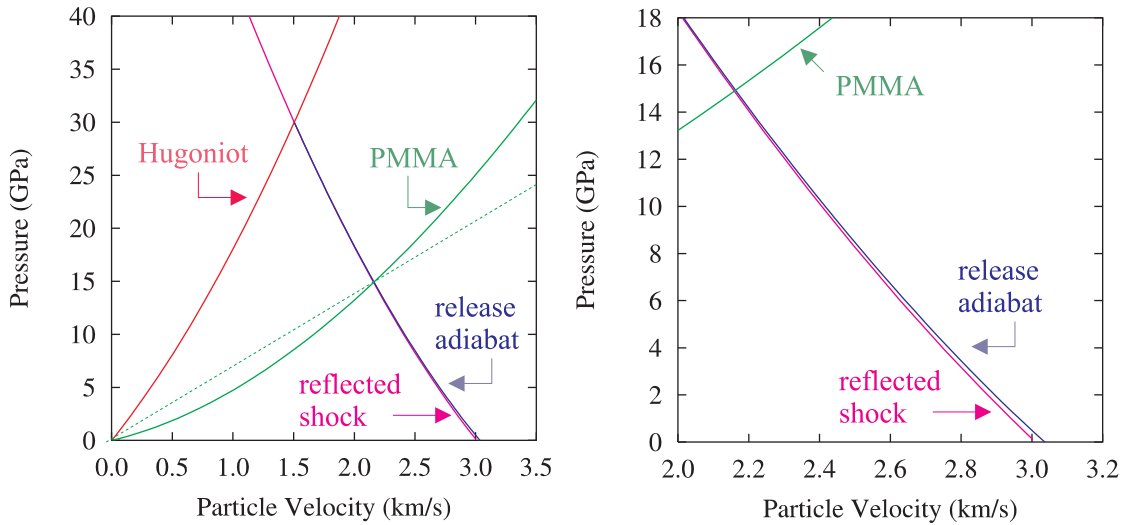

**Fig. 1. Aluminum Hugoniot and release curves in pressure-particle velocity plane. Red—aluminum Hugoniot; blue—aluminum release adiabat from 30 GPa shock state; purple—reflected shock approximation (RSA); green—PMMA Hugoniot; dotted green—line of constant shock velocity for PMMA.**

# INTRODUCTION

The reflected shock approximation to the release adiabat, shown in purple, is a mirror image of the Hugoniot, the mirror plane being a vertical line at particle velocity $u_{P1}$. It is given by the expression

$$P_R(u_R) \approx P_H(2u_{P1} - u_R) \text{ (RSA)}. \qquad (3)$$

Figure 1 shows that the RSA is very close to the true adiabat in this example. The true adiabat gives slightly higher pressures, and the free surface velocity $u_{fs}$ is a factor of 2.017 times the particle velocity—slightly higher than the ideal RSA value of 2.0. The approximation $u_{P1} \approx u_{fs}/2$ would determine the particle velocity to better than 1% in this case.[1]

Now consider an IM experiment in which a standard material is initially in direct contact with a sample material to be measured. This direct contact geometry is especially common in Hugoniot measurements on liquids, where the standard also serves as a sample holder. Direct contact is also used when explosive assemblies are used to generate and/or structure the shock wave in the standard. This method was used in some of the earliest shock wave measurements and is also used in explosive wedge tests. Direct contact also arises in certain cases where the experimentalist does not choose to use the standard material as the impactor.

Suppose that aluminum is used as the standard to measure a point on the Hugoniot for the softer material PMMA, shown by the green curve in Fig. 1. In a typical experiment, one measures the shock velocities $U_S$ of the standard and the sample. The sample particle velocity is then determined from the release adiabat of the standard, by locating the point at which it intersects a line corresponding to the measured $U_S$ for the sample, the curve $P = \rho_0 U_S u_p$, shown by a dotted green line in Fig. 1a. For the case depicted in Fig. 1, using the RSA in place of the exact adiabat would give an error of only 0.2% in the PMMA particle velocity.

The RSA is often used in the analysis of IM experiments because it simplifies the problem. One needs only the Hugoniot of the standard to perform the analysis. Since the Hugoniot is usually available from experimental data, one is not dependent on a theoretical EOS model.

But *why* does the RSA give such good results for aluminum and other materials used as standards in IM experiments? When should one expect it to break down?

Shock waves increase the entropy of the material. Because the adiabat is also an isentrope, the material retains this entropy on unloading. The pressure and temperature on the release adiabat are higher than on the Hugoniot, at the same density, because of the higher entropy. "Conventional wisdom" says that the RSA will give reasonable results if this thermal effect is relatively small, as it is at low pressures. Conversely, if the RSA agrees with the exact adiabat, it is normally assumed that thermal effects are small.

---

1. Corrections to this approximation are sometimes, but not always, taken into account.

## 1.3 Overview of Report

In this report, I will show that the conventional wisdom outlined above is incorrect in several ways. Experience shows that there are many materials for which the RSA is a good enough approximation for the analysis of shock wave experiments (except at very high pressures). However, I will show that this fact is *not* due to the absence of thermal effects. The central points are as follows:

- The arguments about thermal effects do not apply when the Hugoniot and adiabat are mapped from the $P$-$\rho$ plane into the $P$-$u_P$ plane.
- The RSA gives good results *because* of thermal effects, not *in spite* of them. Thermal effects compensate for the mapping effect.
- There are some materials for which the RSA is *not* sufficiently accurate, for which corrections are needed.
- Material strength and porosity can also lead to errors in the RSA.
- Experimental verification of the relation $u_{fs} \approx 2u_P$ does not necessarily prove the accuracy of the RSA for the entire adiabat.

The need for corrections to the RSA at high shock pressures is already well-known (though not always taken into account). But the need for corrections to the RSA at *low* shock pressures, for compressible materials and also materials of high strength, is apparently *not* well known. This report aims to remedy this situation.

Section 2 discusses the simple case in which the Hugoniot and release adiabat are *identical* in the $P$-$\rho$ plane. Using a linear $U_S$-$u_P$ Hugoniot curve, a closed-form solution for the exact adiabat in the $P$-$u_P$ plane is derived and shown to lie *below* the RSA, *not above* it—the opposite of what is expected from thermal effects. Deviations from the RSA are shown to be most important in the more compressible materials. These results are independent of any particular EOS model.

Section 3 compares the RSA with the exact adiabats for three "soft" materials—Teflon, PMMA, and lithium. The largest errors are found in Teflon, where the thermal effects are too small to compensate for the effect of mapping into the $P$-$u_P$ plane. For lithium, where the RSA is quite accurate, the release adiabat is found to have a large thermal offset from the Hugoniot in the $P$-$\rho$ plane.

Section 4 shows that material strength can also cause significant errors in the RSA. The RSA for tungsten, a high-strength material, is shown to lie above the exact adiabat, the same effect as observed when mapping from $P$-$\rho$ into $P$-$u_P$.

Conclusions are given in Sec. 5.

## 2. A SIMPLE CASE

In this section, I will analyze the simple case where the Hugoniot and release adiabat are *identical* in the $P$-$\rho$ plane. For simplicity, I will also assume that the Hugoniot is given by a linear $U_S$-$u_P$ curve. I will demonstrate the following points:

- The RSA does *not* give an exact expression for the release adiabat in the $P$-$u_P$ plane, even in this case where there are no thermal effects.
- The true adiabat lies *below* the RSA in the $P$-$u_P$ plane, the opposite of what is expected from thermal corrections.
- The standard relation, $u_{fs} = 2u_P$, does not hold in this case; $u_{fs} < 2u_P$ for *all values* of the pressure and Hugoniot parameters.
- Corrections to the RSA are most important for materials with low bulk moduli and/or large slopes in the $U_S$-$u_P$ curve.

### 2.1 Basic Equations

Consider a material at zero initial pressure, initial density $\rho_0$, and a linear Hugoniot curve, $U_S = C_0 + S u_P$. For the simple case considered here, the Hugoniot $P_H$ and release adiabat $P_R$ are identical in the $P$-$\rho$ plane and are given by

$$P_H = P_R = \rho_0 C_0^2 \mu / (1 - S\mu)^2, \tag{4}$$

where $\mu$ is the "strain,"

$$\mu = 1 - \rho_0/\rho = u_P/U_S. \tag{5}$$

In the $P$-$u_P$ plane, the Hugoniot is given by

$$P_H = \rho_0 U_S u_P = \rho_0 (C_0 + S u_P) u_P. \tag{6}$$

Consider the release adiabat from a shock state with pressure $P_{H1}$, density $\rho_1$, and particle velocity $u_{P1}$. The rarefaction velocity $u_R$ (particle velocity along the adiabat) is given by Eqs. (1) and (2). Using Eqs. (2) and (5), Eq. (1) becomes

$$u_R = u_{P1} - \int_{\mu_1}^{\mu} \sqrt{\rho_0^{-1} dP_R/d\mu} \, d\mu. \tag{7}$$

## 2.2 Exact Solution for Release Adiabat

The integral in Eq. (7) can be evaluated in closed form. Start by defining the following dimensionless variables:

$$\Phi = SP_R/\rho_0 C_0^2 = S\mu/(1-S\mu)^2 = (1+v)v, \tag{8}$$

$$v = Su_P/C_0, \text{ and} \tag{9}$$

$$\xi = u_R/u_{P1}, \text{ where } \xi_1 = 1. \tag{10}$$

Using these quantities, Eq. (7) becomes

$$\xi = 1 - \frac{C_0}{Su_{P1}} \int_{\mu_1}^{\mu} \sqrt{S\frac{d\Phi}{d\mu}} \, d\mu = 1 - \frac{S}{v_1} \int_{\mu_1}^{\mu} \sqrt{\frac{1+S\mu}{(1-S\mu)^3}} \, d\mu. \tag{11}$$

The strain $\mu$ can be eliminated using the relation $S\mu = v/(1+v)$, which can be obtained from Eq. (8). The result is

$$\xi = 1 - \frac{1}{v_1} \int_{v_1}^{v} \frac{\sqrt{1+2v}}{(1+v)} \, dv = 1 - \frac{2}{v_1}[f(\sqrt{1+2v}) - f(\sqrt{1+2v_1})], \tag{12}$$

where $f(y) = y - \operatorname{atan}(y)$. \hfill (13)

Finally, note that $v$ and $v_1$ can be expressed in terms of $\Phi$ and $\Phi_1$, using Eq. (8),

$$v = \tfrac{1}{2}(\sqrt{1+4\Phi} - 1), \tag{14}$$

and so can be eliminated from the solution. Hence Eqs. (12)-(14) define the function $\xi(\Phi)$ (and thus $\Phi(\xi)$) along the release adiabats for all possible shock states, each corresponding to a different value of $\Phi_1$.

## 2.3 Results

The exact adiabats will now be compared to those obtained from the reflected shock approximation. Using Eqs. (3) and (6), the RSA result, expressed in terms of the reduced variables, is

$$\Phi_{RSA} = \Phi_1 - v_1(1+2v_1)(\xi-1) + v_1^2(\xi-1)^2. \tag{15}$$

Appendix A compares the above quadratic expression with a Taylor series expansion of the exact solution, Eqs. (12)-(14), in powers of $\xi - 1$. It is found that the RSA is not even exact in the leading order terms in the expansion and that the exact solution cannot be truncated after the second-order term.

Figure 2 shows the results for three values of $\Phi_1$, ranging from 0.5 to 2.0. The exact adibat is given in blue, the RSA in red. Figure 3 shows the ratio of the free surface velocity to the particle velocity for shock states with $\Phi_1$ ranging from 0 to 4.

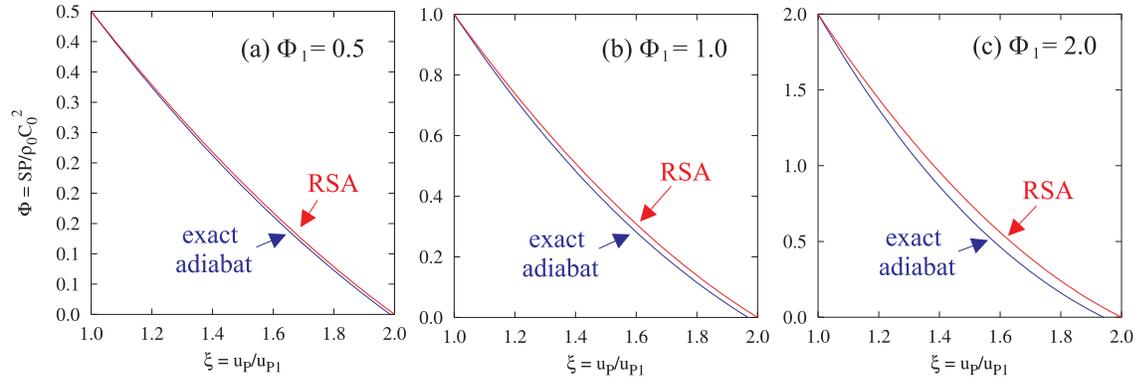

**Fig. 2. Release adiabats for simple case. Exact adiabats, computed from Eqs. (12)-(14), are shown in blue. RSA adiabats, computed from Eq. (15), are shown in red. See the text for discussion of the reduced variables $\Phi$ and $\xi$.**

Figures 2 and 3 show that the RSA does not agree with the exact adiabat for *any* finite value of $\Phi_1$, i.e., that *identity of the Hugoniot and adiabat in the P-$\rho$ plane is not preserved in the P-$u_P$ plane*.

For a given set of Hugoniot parameters ($\rho_0$, $C_0$, $S$), deviations from the RSA become more significant as the shock pressure increases, just as thermal corrections do. However, the exact adiabat lies *below* the RSA—the *opposite* of what occurs when thermal corrections are important. One also finds that $u_{fs} < 2u_{P1}$, *for all* $\Phi_1$—once again, the opposite result from thermal effects.

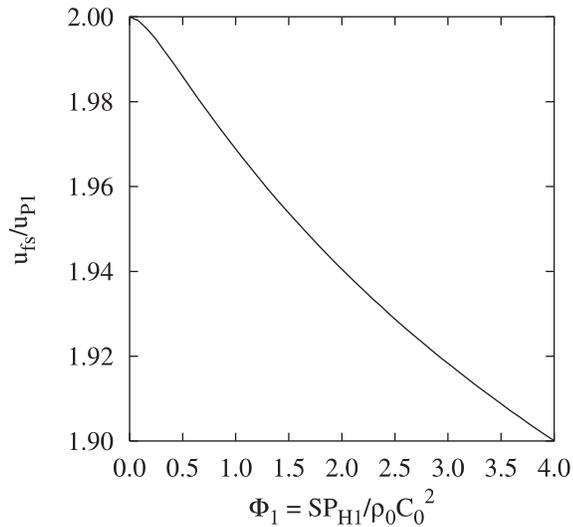

**Fig. 3. Ratio of free surface and particle velocities vs. initial pressure for simple case.**

These results show that the success of the RSA does not arise from the *absence* of thermal effects but rather *because* of them. In fact, thermal corrections actually *offset* the errors in the RSA seen in Figs. 2 and 3—a fortunate, but fortuitous, cancellation of errors. The experimental observation that $u_{fs} \approx 2u_P$ for many materials shows that thermal corrections are *present, not absent*, as often thought.

There is an important corollary to this result: When the RSA agrees with the exact adiabat, the Hugoniot and adiabat *cannot* be identical in the $P$-$\rho$ plane; a large thermal offset must be present.

The RSA is an example of how serendipity has played a role in shock wave physics. The cancellation of errors allows one to analyze IM experiments using only the Hugoniot of the standard, eliminating the need for a theoretical EOS model. However, it should not be assumed that this cancellation occurs in *all* cases. To be on the safe side, one must examine each material to be used as a standard, using a realistic EOS model for the thermal effects and also including corrections for material strength, where necessary.

Figures 2 and 3 show that, apart from thermal and strength effects, corrections to the RSA are important for shock pressures $P_H > 0.5\rho_0 C_0^2/S = 0.5\beta_0/S$. Hence these corrections will be most important for materials having low bulk moduli and/or large slopes of the $U_S$-$u_P$ curve. These conditions are generally found in the more compressible materials, e.g., plastics, polymers, and organic substances, and also the alkali metals. Three cases are discussed in Sec. 3.

In the less compressible materials, including the metals most often used as standards in shock-wave experiments, the errors discussed here are less important because they tend to be offset by thermal effects at high pressures. However, strength effects, which are important in such materials, *can* cause errors similar to those discussed here. Those effects are discussed in Section 4.

## 3. EXAMPLES OF SOFT MATERIALS

In Sec. 2, I showed that the reflected shock approximation (RSA) fails when the release adiabat and the Hugoniot are identical in the $P$-$\rho$ plane, that the identity is not preserved when the adiabat is mapped into the $P$-$u_P$ plane. Hence the success of the RSA arises because thermal effects compensate for this mapping problem. I also showed that these mapping effects are most important for materials having low bulk moduli and/or large slopes of the $U_S$-$u_P$ curve.

In this section, I will discuss two such materials—Teflon and PMMA—where the RSA is inaccurate because the thermal effects do *not* compensate for the mapping effect. I will also show that, for lithium, where the RSA gives good results, there is a large offset between the adiabat and the Hugoniot in the $P$-$\rho$ plane.

Figure 4 compares the exact adiabats (shown in blue) with the RSA adiabats (shown in purple) for Teflon at four shock pressures up to 5 GPa. (Note that Teflon dissociates at ~34 GPa on the Hugoniot, but these calculations correspond to much lower pressures, where there are no dissociation effects.)

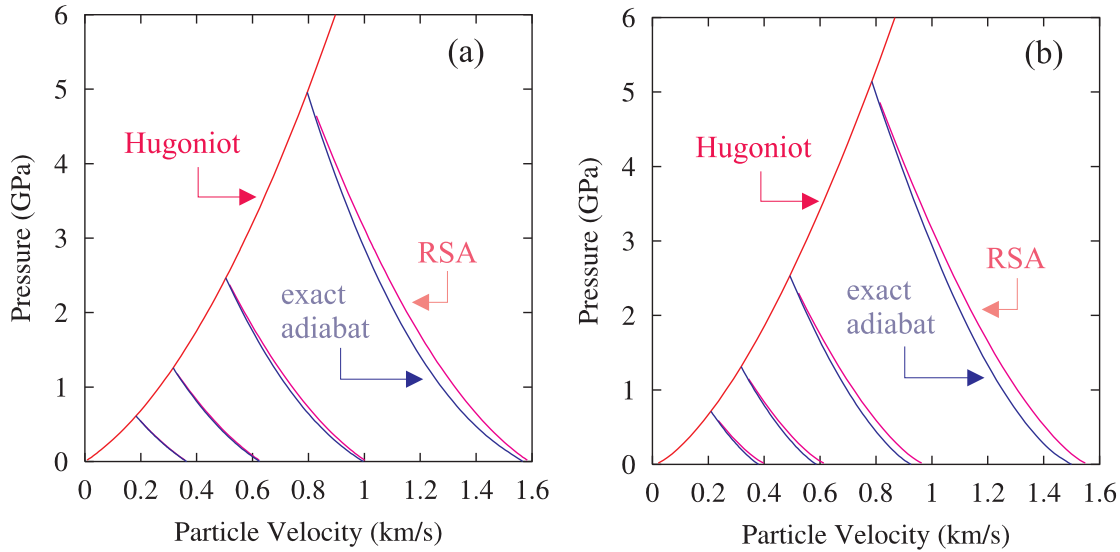

Fig. 4. Exact and RSA adiabats for Teflon. (a)—calculations using a Mie-Grüneisen EOS, (b)—calculations using a tabular EOS and including porosity.

Figure 4a shows the results obtained using a simple Mie-Grüneisen EOS model, fit to experimental data.

$$\rho_0 = 2.15, \ U_S = 1.14 + 2.30 u_P - 0.110 u_P^2, \ \Gamma = 0.50 \rho_0/\rho. \tag{16}$$

Note that the exact adiabat always lies below the RSA curve, just as seen in Sec. 2, the thermal effects being insufficient to compensate for the mapping effect.

Figure 4a also shows that the deviation between the exact and RSA adiabats decreases with decreasing pressure, so that the RSA result, $u_{fs} \approx 2u_P$, is only in error by ~1% (for the 2.5 and 5.0 GPa shock states). This fact shows why *one cannot determine the accuracy of RSA solely by considering the free surface velocity.*

Modeling the behavior of Teflon is complicated by the fact that material samples vary in crystallinity and thus in density [4]. Pressure can transform an amorphous sample to a more crystalline state, with a permanent increase in density. Figure 4b shows the results obtained using a tabular EOS, discussed in Ref. [4], in which this transition is treated using a porosity model. The initial density, 2.15 g/cm$^3$, corresponds to roughly 5% "porosity," which collapses on shocking. The permanent density increase leads to an even greater error in the RSA, and $u_{fs} \approx 2u_P$ is a poor approximation to the free surface velocity. This result shows why *one should not use the reflected shock approximation with porous materials*.

Figure 5 compares the exact adiabats (shown in blue) with the RSA adiabats (shown in purple) for PMMA at four shock pressures up to 20 GPa. (Here again, the shock pressures are below the onset of dissociation at ~25 GPa.) The calculations were made with a Mie-Grüneisen EOS model, discussed in Ref. [5], with

$$\rho_0 = 1.186, \; U_S = 2.30 + 1.75u_P - 0.056u_P^2, \; \Gamma = 0.910\rho_0/\rho. \qquad (17)$$

In this case, deviations between the exact and RSA adiabats are smaller than in Teflon, evidently because of the larger Grüneisen parameter. Notice, however, that the RSA adiabat *crosses* the exact adiabat for the 20 GPa shock pressure. This example shows, once again, that one cannot evaluate the validity of the RSA simply by looking at the free surface velocity. If one were to look solely at $u_{fs}$, one would conclude that the RSA was always a lower bound to the exact adiabat, which is not the case.

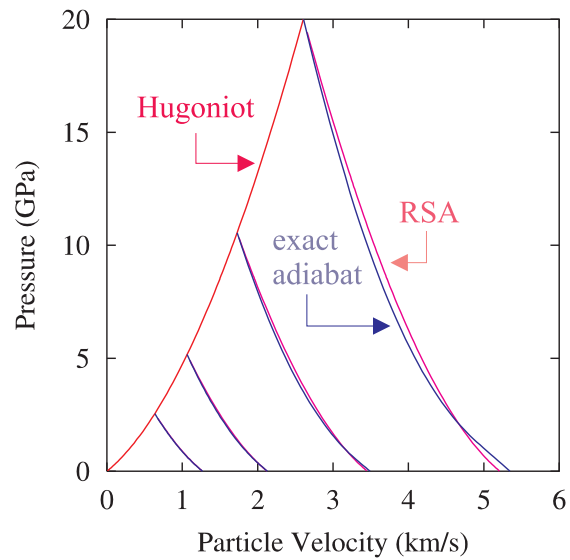

**Fig. 5.** Exact and RSA adiabats for PMMA.

Figure 6a compares the exact adiabats (blue) with the RSA adiabats (purple) for lithium at four shock pressures up to 20 GPa. The calculations were made with a Mie-Grüneisen EOS model, discussed in Ref. [6], with

$$\rho_0 = 0.530, \ U_S = 4.645 + 1.133 u_P, \ \Gamma = 0.810 \rho_0/\rho. \tag{18}$$

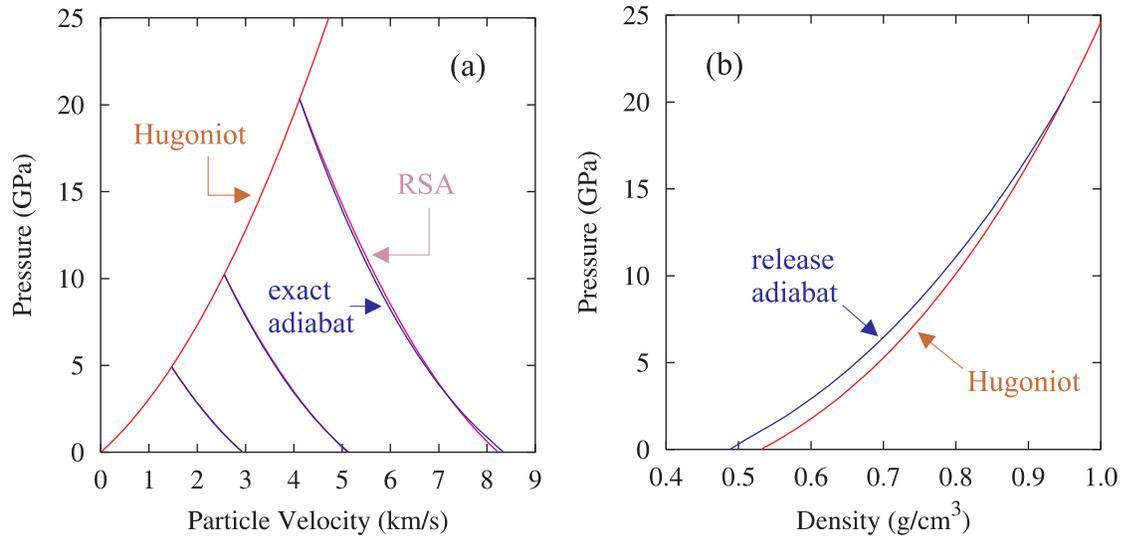

**Fig. 6.** Hugoniot and adiabats for Li. (a)—comparison of exact and RSA adiabats in $P$-$u_P$ plane, (b)—comparison of Hugoniot and 20 GPa release adiabat in $P$-$\rho$ plane.

In this case, the RSA is a *good* approximation to the exact adiabat. The reason for the good agreement can be seen in Fig. 6b, which compares the Hugoniot and the adiabat for the 20 GPa shock state in the $P$-$\rho$ plane. There is a large thermal offset between the Hugoniot and adiabat; it is precisely this offset that accounts for the good agreement in the $P$-$u_P$ plane. Once again, we see that *the success of the RSA does not arise from the lack of thermal effects—it arises because of thermal effects*, which compensate for the mapping effect.

# 4. MATERIAL STRENGTH EFFECTS

In Secs. 2 and 3, I showed that the reflected shock approximation (RSA) is not accurate for certain soft materials, in which thermal effects do not compensate for the effect of mapping the release adiabat from the $P$-$\rho$ plane to the $P$-$u_P$ plane. However, those materials are not generally used as standards in impedance matching experiments.

In this section, I will discuss a completely different issue that could lead to errors in the RSA for *hard* materials, i.e., the effects of material strength.

In order to demonstrate these effects, I have deliberately chosen a high-strength material—tungsten—and have assumed an idealized elastic-perfectly plastic model. I have used a tabular EOS, taken from Ref. [7], along with strength parameters from Steinberg's tabulation [8]. The release adiabats were first computed in the $P$-$\rho$ plane and then mapped into the $P$-$u_P$ plane, using Eq. (1). The calculations were carried out using the EOSPro code [3].

Figure 7a shows the Hugoniot, with and without the strength terms, and a release adiabat from a 50 GPa shock state, in the $P$-$\rho$ plane[1]. On release, the material first unloads elastically, with a sharp drop in stress to ~38 GPa. At lower stresses, the adiabat is roughly parallel to the Hugoniot.

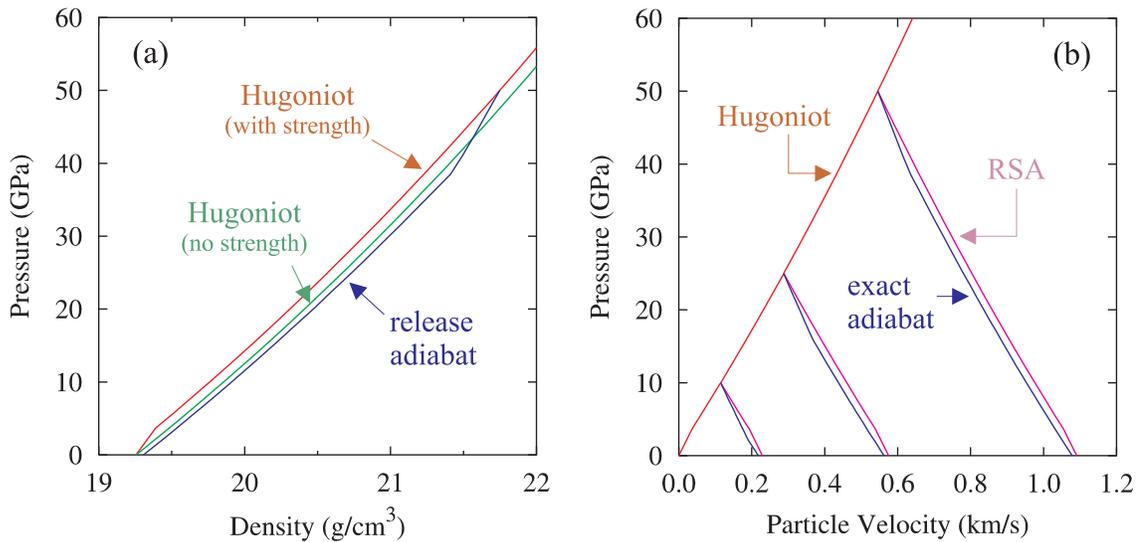

**Fig. 7.** Effect of strength on release adiabats for tungsten. (a)—comparison of Hugoniot and 50 GPa release adiabat in $P$-$\rho$ plane; (b)—comparison of exact and RSA adiabats in $P$-$u_P$ plane.

---

1. Strictly speaking, I am plotting stress, not pressure.

Figure 7b shows the Hugoniot and three release adiabats in the $P$-$u_P$ plane. Here again, the RSA adiabats lie above the exact adiabats, just as was seen in Secs. 2 and 3. The RSA is in error by as much as 5%; that would be a *significant* source of error in an IM experiment.

Understanding the effect of material strength on impedance matching experiments presents real challenges. In particular, the validity of the elastic-perfectly plastic model of release behavior is not known, particularly for shocks at very high pressures. Exploring this issue in depth would require well thought out experiments as well as theoretical analysis. However, the above example indicates that further investigation is needed.

## 5. CONCLUSIONS

In this report, I have shown that the success of the reflected shock approximation (RSA) arises from a fortunate, but fortuitous cancellation of errors. In the absence of a thermal offset between the Hugoniot and the release adiabat, in the $P$-$\rho$ plane, the release adiabat will always lie below the RSA, in the $P$-$u_P$ plane. Porosity and material strength will also cause the adiabat to lie below the RSA. In most materials, thermal effects compensate for these effects so that the RSA turns out to be a fairly good approximation to the exact adiabat, over a surprisingly wide range of pressures.

However, this cancellation does not occur in all materials and should not be regarded as a "theorem" or universal result. Section 3 gives examples of two "soft" materials, Teflon and PMMA, in which the cancellation does *not* occur, even at very low shock pressures. Section 4 gives an example of a high-strength material, tungsten, in which material strength effects lead to significant errors in the RSA.

The wide-spread success of the RSA demonstrates that thermal effects are important in describing release behavior, not unimportant, as often thought. Therefore, one should not assume that an EOS model is "good enough" for practical applications simply because it matches the experimental Hugoniot data and the shock pressures are relatively low.

The fact that errors in the RSA can occur at *low* shock pressures, not just at high pressures, apparently is not well-known in the shock-wave community. I have not found any discussion of this matter in the literature. I would be grateful to know of any earlier publications on the subject.

Of course, it is already known that additional errors in the RSA can occur at high pressures, where thermal effects become very large, due to melting, chemical reactions, thermal electronic excitation, etc. I have not discussed those effects in this report. However, I will conclude by mentioning two experimental papers, [9] and [10], in which my EOS models for Al and Pb have been found to give good agreement with measurements of the release curves at very high shock pressures.[1]

---

1. In Ref. [10], the authors claim that my theoretical EOS for Pb includes the effects of thermal ionization but not pressure ionization. That is incorrect. Pressure ionization is included in *all* my EOS models, for *all* materials.

# REFERENCES

[1] M. H. Rice, R. G. McQueen, and J. M. Walsh, "Compression of Solids by Strong Shock Waves," Solid State Phys. 6, 1-63 (1958).

[2] G. I. Kerley, "Theoretical Equation of State for Aluminum," Int. J. Impact Engng. 5, 441-449 (1987).

[3] EOSPro is a new version of the PANDA code, which is described in: G. I. Kerley, "User's Manual for PANDA II: A Computer Code for Calculating Equations of State," Sandia National Laboratories report SAND88-2291, 1991.

[4] G. I. Kerley, "A Reactive Equation of State Model for Teflon," Kerley Publishing Services report KPS96-10, Albuquerque, NM, October 1996.

[5] G. I. Kerley, "Equations of State for Composite Materials," Kerley Publishing Services report KPS99-4, Albuquerque, NM, December 1999.

[6] G. I. Kerley, "CTH Equation of State Database: MGRUN Option," Kerley Publishing Services report KPS00-4, Albuquerque, NM, June 2000.

[7] G. I. Kerley, "Equations of State for Be, Ni, W, and Au," Sandia National Laboratories report SAND2003-3784 (2003).

[8] D. J. Steinberg, "Equation of State and Strength Properties of Selected Materials," Lawrence Livermore National Laboratory report UCRL-MA-106439, 1991.

[9] M. D. Knudson, J. R. Asay, and C. Deeney, "Adiabatic release measurements in aluminum from 240- to 500-GPa states on the principal Hugoniot," J. Appl. Phys. 97, 073514;1-14 (2005).

[10] S. D. Rothman, K. Parker, C. Robinson, and M. D. Knudson, "Measurement of a release adiabat from ~8 Mbar in lead using magnetically driven flyer impact," Physics of Plasmas 11, 5620-5625 (2004). See footnote on pg. 19.

# Appendix A

# Series Approximation for Simple Case

In Sec. 2, I considered a simple case in which the Hugoniot and adiabat are identical in the $P$-$\rho$ plane and the Hugoniot is given by a linear $U_S$-$u_P$ curve. The exact solution for the adiabat, in the $P$-$u_P$ plane, is given by Eqs. (12)-(14). The reflected shock approximation to the adiabat is given by Eq. (15), which can be written

$$\Phi_{RSA} = \Phi_1 + C_1(\xi - 1) + C_2(\xi - 1)^2, \tag{A.1}$$

where $\Phi$, $\nu$, and $\xi$ are reduced variables, defined by Eqs. (8)-(10), and

$$C_1 = -\nu_1(1 + 2\nu_1), \quad C_2 = \nu_1^2. \tag{A.2}$$

In this appendix, I will expand the exact solution in a Taylor series about $\xi - 1$ and compare the coefficients with the quadratic expression, Eq. (A.1), for the RSA. Begin with

$$\Phi(\xi) = \Phi_1 + \sum_{k=1}^{\infty} A_k (\xi - 1)^k, \tag{A.3}$$

where

$$A_k = \frac{1}{k!}(d^k \Phi / d\xi^k)_{\xi = 1}. \tag{A.4}$$

After carrying out the differentiations, the first four coefficients are found to be

$$A_1 = -\nu_1(1 + \nu_1)\sqrt{1 + 2\nu_1}, \tag{A.5}$$

$$A_2 = \nu_1^2(1 + \nu_1)(1 + \tfrac{3}{2}\nu_1)/(1 + 2\nu_1), \tag{A.6}$$

$$A_3 = -\tfrac{1}{6}\nu_1^3(1 + \nu_1)(1 + 6\nu_1 + 6\nu_1^2)/(1 + 2\nu_1)^{5/2}, \text{ and} \tag{A.7}$$

$$A_4 = \tfrac{1}{24}\nu_1^4(1 + \nu_1)(2 + 3\nu_1 + 6\nu_1^2 + 6\nu_1^3)/(1 + 2\nu_1)^4. \tag{A.8}$$

Hence the exact adiabat is clearly quite different from the RSA. The following relationship holds between the first- and second-order coefficients:

$$A_1/C_1 = (1+v_1)/(\sqrt{1+2v_1}) \geq 1, \tag{A.9}$$

$$A_2/C_2 = (1+v_1)(1+\tfrac{3}{2}v_1)/(1+2v_1) \geq 1. \tag{A.10}$$

Hence the RSA expression is not even exact in the leading order terms in the expansion. Moreover, the series for exact adiabat is not accurate when truncated after the quadratic term. (The fourth-order series is very accurate for the results shown in Fig. 2. Higher-order terms are needed for larger values of $\Phi_1$.)